\documentclass[12pt]{article}
\usepackage{hyperref}
\usepackage{scalerel}
\usepackage[normalem]{ulem}
\usepackage{amsfonts}
\usepackage{bm}
\usepackage{cite}
\usepackage{mathrsfs}
\usepackage{amsmath}
\usepackage{graphicx}
\usepackage{hyperref}
\usepackage{verbatim}
\usepackage{color}
\usepackage[dvipsnames]{xcolor}
\usepackage{enumerate}
\usepackage{amsfonts}
\usepackage{amssymb}
\usepackage{authblk}
\usepackage[hang,flushmargin]{footmisc}
\usepackage{lipsum}
\usepackage[font=footnotesize]{caption}
\usepackage{soul}
\DeclareMathAlphabet\mathbfcal{OMS}{cmsy}{b}{n}
\usepackage{empheq}
\usepackage{cancel}

\csname @addtoreset\endcsname{equation}{section}
\newcommand{\eal}[1]{\begin{equation} \begin{aligned} #1 \end{aligned}\end{equation}}

 \newcommand{\badat}{\begin{alignedat}}
 \newcommand{\eadat}{\end{alignedat}}

\def\be{\begin{eqnarray}}
\def\ee{\end{eqnarray}}
\def\beann{\begin{eqnarray*}}
\def\eeann{\end{eqnarray*}}
\def\beq{\begin{equation}}
\def\eeq{\end{equation}}
\def\ba{\begin{array}}
\def\ea{\end{array}}
\def\ben{\begin{enumerate}}
\def\een{\end{enumerate}}
\def\bea{\begin{eqnarray}}
\def\eea{\end{eqnarray}}

\def\5{\bar }
\def\6{\partial }
\def\7{\hat }
\def\4{\tilde }
\renewcommand{\d}{\partial}

\def\cA{\mathcal{A}}

\def\cD{\mathcal{D}}
\def\cE{\mathcal{E}}
\def\cF{\mathcal{F}}
\def\cG{\mathcal{G}}
\def\cH{\mathcal{H}}

\def\cJ{\mathcal{J}}

\def\cM{\mathcal{M}}
\def\cN{\mathcal{N}}
\def\cO{\mathcal{O}}
\def\cP{\mathcal{P}}

\def\cV{\mathcal{V}}

\def\boe{\boldsymbol{e}}
\def\bl{\boldsymbol{\tau}}
\def\bT{{\bf T}}

\def\bZ{{\bf Z}}
\def\bGa{{\bf \Gamma}}

\hypersetup{final}

\begin{document}
\begin{titlepage}
  \thispagestyle{empty}
  
  \begin{center}
	 \vskip2cm
  \baselineskip=14pt {\LARGE \scshape{The structure of IR divergences\\[10pt] in celestial gluon amplitudes}}\\ 
   \vskip1.5cm 
   %\today
	% \vskip1.5cm
   \centerline{ 
   {Hern\'an A. Gonz\'alez${}^a$  and}
   {Francisco Rojas}${}^b$
   }

 \bigskip\bigskip
 
 \centerline{\em${}^a$\small Departamento de Ciencias, Facultad de Artes Liberales, Universidad Adolfo Ib\'a\~nez,  Chile}
 
\smallskip
 
\centerline{\em${}^b$\small Facultad de Ingenier\'ia y Ciencias, Universidad Adolfo Ib\'a\~nez, Chile}

\bigskip\bigskip
  
  \end{center}

\begin{abstract}
\noindent The all-loop resummation of SU$(N)$ gauge theory amplitudes is known to factorize into an IR-divergent (soft and collinear) factor and a finite (hard) piece. The divergent factor is universal, whereas the hard function is a process-dependent quantity. 

 We prove that this factorization persists for the corresponding celestial amplitudes. Moreover, the soft/collinear factor becomes a scalar correlator of the product of renormalized Wilson lines defined in terms of celestial data. Their effect on the hard amplitude is a shift in the scaling dimensions by an infinite amount, proportional to the cusp anomalous dimension. This leads us to conclude that the celestial-IR-safe gluon amplitude corresponds to a expectation value of operators dressed with \emph{Wilson line primaries}. These results hold for finite $N$.

In the large $N$ limit, we show that the soft/collinear correlator can be described in terms of vertex operators in a Coulomb gas of colored scalar primaries with nearest neighbor interactions. In the particular cases of four and five gluons in planar $\cN=4$ SYM theory, where the hard factor is known to exponentiate, we establish that the Mellin transform converges in the UV thanks to the fact that the cusp anomalous dimension is a positive quantity. In other words, the very existence of the full celestial amplitude is owed to the positivity of the cusp anomalous dimension.

\end{abstract}

 \end{titlepage}
\tableofcontents

\section{Introduction}

Infrared (IR) divergences are ubiquitous in gauge theory scattering amplitudes. They arise when charged external states interact with an infinite amount of virtual excitations with arbitrarily low energies. The existence of these infinities suggests that the usual Fock space, constructed out of a unique vacuum, is not properly defined. Consequently, the notions of asymptotic one-particle states, and the vacuum itself, need to be revised.

\vspace{2mm}

Despite these issues, the most common strategy to deal with IR singularities is to focus on physical quantities such as the total cross section. Indeed, it was proven that IR-safe observables can be extracted from QED \cite{Bloch:1937pw} and QCD  \cite{Lee:1964is,Kinoshita:1975bt} cross sections, by a careful cancellation between virtual and real emissions, order-by-order in perturbation theory. 

\vspace{2mm}

Nonetheless, there is another way of dealing with infinities arising from these long-range interactions, namely, by extracting the IR-safe information from the amplitude itself. Due to the universal structure of the infrared radiation, it is possible to separate the different energy scales involved in the collision. Using the eikonal approximation, one chooses a scale $Q$ where the finite contribution to the amplitude is consistently separated from the soft and collinear singularities. This type of organization in QCD has been the focus of extensive work \cite{Collins:1980ih,Sen:1981sd,Sen:1982bt,Collins:1989bt,Kidonakis:1998nf,Sterman:1995fz,Feige:2014wja}.

\vspace{2mm}

The key idea behind the IR factorization of gauge theory amplitudes is that, since the energy scale is arbitrary,  soft factors must be independent of $Q$. Then, one can use renormalization group techniques to treat the singular piece in terms RG flow equations. This approach imposes strong constraints on the IR structure of the amplitudes, resulting in exponentiated expressions valid at all orders in perturbation theory \cite{Dixon:2008gr,Gardi:2009qi,Gardi:2009zv,Becher:2009cu,Becher:2009qa}. 

\vspace{2mm}

Similarly, a deep connection has been shown to exist between IR divergences in amplitudes and UV divergences in Wilson line correlators \cite{Polyakov:1980ca,Brandt:1981kf,Frenkel:1984pz,Gardi:2013ita}, permitting a gauge invariant characterization of these singularities. In particular, it is worth highlighting the work of Korchemsky and Radyushkin \cite{Korchemsky:1985xj,Korchemsky:1987wg}, who showed that the renormalization properties of cusped light-like Wilson loops universally describe the pole structure of the regularized gauge theory $S$-matrix. The quantity controlling these poles is known as the \emph{cusp anomalous dimension}, $\gamma(\alpha)$, a perturbative function of the gauge theory coupling $\alpha$.

\vspace{2mm}

Over the last years, there has been a renewed interest in understanding the role of IR divergences in gauge theory and gravity from a perspective based on symmetries \cite{Strominger:2013jfa,Strominger:2013lka,He:2014laa,Campiglia:2014yka,He:2015zea,Kapec:2017tkm,Choi:2017ylo,Strominger:2017zoo,Hirai:2019gio,He:2020ifr,Hirai:2020kzx}. The central point in this discussion is that the degenerated vacuum, observed in theories with massless excitations, is greatly explained by the presence of an infinite set of conserved charges. 

\vspace{2mm}

These quantities correspond to surface integrals located at the null boundary of the spacetime that were first discovered in \cite{Bondi:1962px,Sachs:1962wk} and further developed in a holographic context in \cite{Barnich:2010eb,Strominger:2013lka}. Their importance in this discussion is twofold: on the one hand they do not change the energy of the process, thus inducing an infinite degeneracy of the vacuum, but introduce degrees of freedom helping to account for the universal organization of IR radiation; on the other hand, their two-dimensional nature shows an intriguing holographic realization of generic 4D gauge theories.  

\vspace{2mm}

Along these lines, celestial amplitudes provide an specific realization of a holographic description of flat spacetime physics. They are a map between 4D scattering amplitudes for massless particles and conformal correlators on the sphere at null infinity \cite{Cheung:2016iub,Pasterski:2017ylz}, known as the celestial sphere. The isomorphism between the Lorentz group and ${\rm SL}(2,\mathbb{C})$ identifies the usual plane-wave asymptotic states with conformal primaries on the celestial sphere whose CFT correlators are built upon \cite{Pasterski:2017kqt}. Progress along these lines encompassing different perspectives in these topics can be found in \cite{Donnay:2018neh,Schreiber:2017jsr,Stieberger:2018edy,Stieberger:2018onx,Fan:2019emx,Fotopoulos:2019vac,Fotopoulos:2019tpe,Fan:2020xjj,Fotopoulos:2020bqj,Pate:2019mfs,Pate:2019lpp,Puhm:2019zbl,Adamo:2019ipt,Nandan:2019jas,Guevara:2019ypd,Law:2019glh,Albayrak:2020saa,Gonzalez:2020tpi,Casali:2020vuy,Casali:2020uvr,Banerjee:2020kaa,Banerjee:2020zlg,Banerjee:2020vnt,Donnay:2020guq,Banerjee:2021cly,Narayanan:2020amh,Kalyanapuram:2020epb,Pasterski:2020pdk,Guevara:2021abz,Barnich:2021dta}.

Two-dimensional correlators are obtained through a Mellin transform of the bulk amplitude. More concretely, the Mellin transform involves an integration over all the energies of the external particles participating in the scattering process, from the deep infrared to the  ultraviolet. Hence, their study may reveal yet another aspect of the UV/IR connection, which could help to further constrain the $S$-matrix. In other words, they may impose a new kind of high energy consistency relations onto scattering amplitudes.

\vspace{2mm}

In this article, we connect the singular behavior exhibited by all-loop IR regulated $n$-point gluon amplitudes with 2D conformal field theory structures. We follow and extend the approach recently presented in \cite{Arkani-Hamed:2020gyp}, showing that the factorization of divergences in non-abelian gauge theories \cite{Polyakov:1980ca,Brandt:1981kf,Frenkel:1984pz,Korchemsky:1985xj,Korchemsky:1987wg,Gardi:2009qi,Becher:2009cu,Gardi:2013ita,Becher:2019avh} persists in a universal fashion when expressed in the celestial basis. More concretely, we first identify the hard contribution as a celestial correlator with shifted conformal dimensions. This shift corresponds to an integral of the cusp anomalous dimension over the energy scales involved. This quantity becomes divergent once the IR-regulator is removed. Furthermore, we notice that the divergent factor, consisting of soft and collinear singularities, can be expressed as a correlator of colored primary fields on the celestial sphere, given by Wilson lines anchored to each external particle. By carefully analyzing the SL$(2,\mathbb{C})$ covariant properties of these Wilson line operators, we find that their conformal weights have the exact same shift observed in the hard correlator. This leads us to conclude that the celestial-IR-safe gluon amplitude corresponds to a correlator of operators dressed with Wilson line primaries.  

\vspace{2mm}

The organization of this paper is as follows. In section \ref{Sec2}, using the BDS formula, we examine the celestial factorization of gluon amplitudes in the large $N$ limit of maximally supersymmetric Yang-Mills \cite{Bern:2005iz}. In \ref{divlargeN}, the divergent piece is shown to be effectively described by a Coulomb gas of scalar colored primaries with nearest neighbour interactions. Moreover, in subsection \ref{fin}, it is shown that for four and five gluons, given that the hard factor of these amplitudes also exponentiates, the convergence of the Mellin transform in the UV is ensured by the positivity of the cusp anomalous dimension. 

\vspace{2mm}

Section \ref{Sec3} generalizes the results of the previous section to SU$(N)$ non-abelian gauge theory. We start by describing, using the color-space formalism \cite{Catani:1996jh}, generic properties of all-loop IR regulated gluon amplitudes in momentum basis, based on \cite{Gardi:2009qi,Becher:2009cu,Becher:2019avh}. Subsection \ref{Wilson} is devoted to describe IR divergences in terms of correlators of Wilson lines. In \ref{celestialQCD}, we show that the hard celestial correlator is built out of operators with infinitely shifted conformal dimensions as shown in \eqref{shiftQCD}. At the end of this subsection, it is shown that this exact same shift arises as the anomalous dimension of renormalized Wilson lines. It is then argued that the anomalous dimension corresponds to the conformal weight of these operators. We conclude this section in \ref{LargeNQCD}, where the large $N$ limit of SU$(N)$ celestial gluon amplitudes is discussed, and we demonstrate that the structure of the divergences matches \eqref{divergent-planar}. 
We close this article with section \ref{sec:remarks}, analyzing possible corrections to the results presented here and describing future work. Appendix A is devoted to show that, after a simple observation, the full expression for the celestial BDS formula of \cite{Gonzalez:2020tpi} it is directly mapped to our results of section \eqref{celestialBDS}.

\vspace{6mm}

\noindent \emph{Note added:} While this paper was being finalized, overlapping results appeared in \cite{Magnea:2021tt}.

\section{Celestial amplitudes in \texorpdfstring{$\mathcal{N}=4$}{N=4} SYM}
\label{Sec2}
\subsection{Review of the BDS Ansatz}
\label{s2BDS}
We review known results of all-loop $n$-gluon amplitudes in maximally supersymmetric SU$(N)$ Yang-Mills theory (SYM), for maximal helicity violation (MHV). Perturbatively, they can be expressed as
\be
\label{sumcM}
\cM_{n}=\sum^{\infty}_{L=0} \alpha^{L} \cA^{(L)}\,\delta\left(\sum^{n}_{i=1} p^\mu_i\right)\,,
\ee
with the Dirac delta enforcing momentum conservation and  $\alpha= 2 e^{-\epsilon \gamma_E} g^2/(4\pi)^{2-\epsilon}$ being the Yang-Mills coupling constant. The term $\cA^{(L)}$ is the $L$-loop contribution that is regularized by computing momentum integrals in  $D=4-2\epsilon$ dimensions. The amplitude $\cA^{(L)}$ is decomposed in single and multi-traces of generators in the fundamental representation of SU$(N)$. In the large $N$ limit, the leading contribution is controlled by single traces of $n$ generators and reads
\be
\label{AL}
\cA^{(L)}= g^{n-2} N^L \sum_{\pi} {\rm tr}\left[t^{a_{\pi(1)}}\cdots t^{a_{\pi(n)}} \right] A^{(L)}\left[\pi(1),\cdots,\pi(n)\right]\,,
\ee
where $\pi$ runs over non-cyclic permutations of the external legs. The expression above helps us disentangle the color structure (given by the trace factors) from the dynamical content. This property can be used to examine the kinematics of the all-loop planar gluon amplitude with a specific color ordering. Hence, in the rest of this section, we focus on the ordering
\be
\label{sumM}
M_{n}\equiv g^{n-2} \sum^{\infty}_{L=0} a^{L} A^{(L)}[1,2,\cdots,n]\,\delta\left(\sum^{n}_{i=1} p^\mu_i\right) ,\quad a=\frac{N \alpha}{2\pi} \,(4\pi e^{-\gamma_E})^{\epsilon}\,,
\ee
with the 't-Hooft coupling $a$ beign now the perturbative expansion parameter. In 2005, Bern, Dixon and Smirnov \cite{Bern:2005iz}, based on previous work \cite{Anastasiou:2003kj}, were able to show that the sum \eqref{sumM} actually exponentiates to
\be
\label{BDS_TY}
\sum^{\infty}_{L=0} a^{L} A^{(L)}=A_{\rm tree} \, \exp \left[ -\frac{1}{2}\sum_{l\geq1} a^l \left(\frac{\gamma^{(l)}}{4 (l \epsilon)^2} +  \frac{\cG^{(l)}}{2 l \epsilon} \right) \sum^{n}_{i=1} \left(-\mu^2 s_{i,i+1}^{-1}\right)^{l \epsilon} 
+F_{n}(a,s_{i,j})
+O(\epsilon)\right] \,,
\ee
where $A_{\rm tree}$ is the color-ordered MHV tree-level amplitude \cite{Parke:1986gb} containing all the polarization-dependent information, while the exponential factor carries the full infrared-divergent structure as well as a finite contribution. Kinematically, the exponent above depends only on the Mandelstam invariants $s_{i,j}=2 p_i \cdot p_j$, with the identification $s_{n,n+1}\equiv s_{n,1}$, where $p_i$ is the momentum of each external gluon.\footnote{We work with the metric $\eta={\rm diag}(+,-,-,-)$} It is interesting to note that, due to the planar limit, interactions among neighboring external particles are the only ones contributing with the divergent terms in the expression above. We will investigate the implications of this organization in the celestial basis in subsection \ref{divlargeN}.

Since $\cN=4$ super Yang-Mills is a UV-finite theory \cite{Brink:1982wv,Howe:1983sr}, the remaining divergences are soft and collinear singularities originating from integrations over internal loops. In order to regulate these divergences one needs $\epsilon<0$, a condition that will be used throughout this article.

In expression \eqref{BDS_TY}, double and single poles in $\epsilon$ are controlled by the coefficients $\gamma^{(l)}$ and $\cG^{(l)}$, consisting in the $l$-loop contributions to the cusp and collinear anomalous dimensions respectively, \emph{i.e.},
\be
\label{gaG}
\gamma(a)=\sum_{l\geq 1} a^l \gamma^{(l)}\,, \quad \cG(a)=\sum_{l\geq 1} a^l \cG^{(l)}\,.
\ee
These quantities arise as coefficients in the renormalization group equations for certain observables, such as Wilson loops and form factors. The cusp anomalous dimension $\gamma(a)$ shows up in the renormalization of the product of two semi-infinite Wilson lines \cite{Polyakov:1980ca,Brandt:1981kf,Korchemsky:1987wg,Korchemsky:1991zp}, while $\cG(a)$ makes its appearance in the RG flow that defines the Sudakov form factor \cite{Magnea:1990zb}.

What makes the formula \eqref{BDS_TY} particularly appealing is the appearance of the full finite part of the amplitude, denoted by $F_n$. For $n=4$ and $n=5$, $F_{n}$ is directly expressible in terms of one-loop data, however, starting at $n=6$, it also acquires higher loop corrections depending on dual conformal invariant cross-ratios of the Mandelstam variables \cite{Drummond:2007cf,Drummond:2007au,Bern:2008ap,Cachazo:2008hp}.

In what follows, we will make use of the celestial basis \cite{Pasterski:2017kqt} to interpret the role played by the divergences in \eqref{BDS_TY} as the vacuum expectation value of operators in a two-dimensional CFT. Furthermore, these operators will become a crucial ingredient when extracting the infrared-safe information from \eqref{BDS_TY} and translating it to the celestial sphere. 
%%%%%%%%%%%%%%%%%%%%%%%%%%%%%%%%%%%%%%%%%%%%%%%%%%%%%%%%%%%%%%
\subsection{IR divergences in celestial \texorpdfstring{$n$}{n}-point correlators}
\label{infh-s}
%%%%%%%%%%%%%%%%%%%%%%%%%%%%%%%%%%%%%%%%%%%%%%%%%%%%%%%%%%%%%%
The momentum of each external massless particle in an scattering process will be parameterized in terms of complex coordinates $(z_i,\bar{z}_i)$ and a real number $\omega_i$ as
\be
p_i^{\mu}=\eta_i \omega_i q^\mu(z_i,\bar{z}_i)=\frac{1}{2}\eta_i \omega_i (1+|z_i|^2,z_i+\bar{z}_i, -i(z_i-\bar{z}_i), 1-|z_i|^2)\,,
\ee
where $\eta_i$ is $\pm1$ for outgoing/incoming particles. In terms of the above decomposition, the Mandelstam variables read
\be
\label{mande}
s_{ij}=\eta_i \eta_j \omega_i \omega_j \left|z_{ij}\right|^2\,.
\ee
with $z_{ij} = z_i-z_j$. We now consider the Mellin transform on each external leg of the BDS amplitude
\be
\label{MBDS}
\tilde \cM_{n}\left( \left\{z_i,\bar{z}_i,\Delta_i\right\} \right) = \prod_{i=1}^n  \left(\int_0^\infty d\omega_i \, \omega_i^{\Delta_i-1}\right) M^{\rm BDS}_n\left( \left\{z_i,\bar{z}_i,\omega_i\right\} \right)\,.
\ee
As it has been shown in \cite{Cheung:2016iub,Pasterski:2017kqt,Pasterski:2017ylz}, the main property of a Mellin transformed amplitude is its covariance under SL$(2,\mathbb{C})$ transformations. Since this corresponds to the global part of the conformal group in two dimensions, one then expects that \eqref{MBDS} can be expressed as a correlator of $n$ insertions on the celestial sphere 
\be
\label{cecorr}
\tilde \cM_{n}\left( \left\{z_i,\bar{z}_i,\Delta_i\right\} \right) = \langle \cO_{\Delta_1,\ell_1}(z_1,\bar{z}_1) \cdots \cO_{\Delta_n,\ell_n}(z_n,\bar{z}_n)\rangle\,,
\ee
where one can identify the conformal primary $\cO_{\Delta_i,\ell_i}(z_i,\bar{z}_i)$, having weights $h_i=\tfrac{1}{2}(\Delta_i+\ell_i)$ and $\bar{h}_i=\tfrac{1}{2}(\Delta_i-\ell_i)$, with $\ell_i$ being the helicity of each external particle. Note that for the MHV amplitudes considered here, all but two of the spins $\ell_i$ are positive.

Whenever infrared divergences are present in an amplitude, it is clear that \eqref{cecorr} is not a well-defined object as it will inherit these singularities when taking $\epsilon\to0$. However, in order to define a celestial correlator free of divergences, we will show that soft and hard gluonic degrees of freedom decouple in the celestial basis. In fact, it has been observed in \cite{Arkani-Hamed:2020gyp} that for gravity and scalar electrodynamics, celestial amplitudes can be arranged to be written as
\be
\label{S-H}
\tilde \cM_{n}\left( \left\{z_i,\bar{z}_i,\Delta_i\right\} \right) = \langle\cV_{\sigma_1} (z_1,\bar{z}_1)  \cdots \cV_{\sigma_n} (z_n,\bar{z}_n)  \rangle \langle \hat{\cO}_{\Delta_1,\ell_1}(z_1,\bar{z}_1)  \cdots \hat{\cO}_{\Delta_n,\ell_n}(z_n,\bar{z}_n)\rangle\,,
\ee
where the first  factor controls the IR (soft) divergences through a correlator of $n$ operators $\cV_{\sigma_i}(z_i,\bar{z}_i)$. These are primaries with zero spin and conformal dimension $2\sigma_i$ that are both divergent and regulator-dependent. 

Based on this factorization, one defines the infrared-safe part of the celestial amplitude as the second factor in \eqref{S-H}. This corresponds to the expectation value of $n$ dressed operators $\hat{\cO}_{\Delta_i,\ell_i}$ with conformal dimension $\Delta_i -\sigma_i$. It is then appealing to suggest, for gluon amplitudes, the normal-ordered relation
\be
\label{dressed0}
\cO_{\Delta,\ell}(z,\bar z)=:\cV_{\sigma}(z,\bar z) \, \hat{\cO}_{\Delta,\ell}(z,\bar z):\,,
\ee
that has, in fact, already appeared in \cite{Nande:2017dba,Himwich:2020rro} for electrons and gravitons, respectively. In the next section we focus on the factorization \eqref{S-H} and we reveal the role of \eqref{dressed0} in the case of celestial BDS amplitudes.

\subsection{BDS celestial amplitude}
\label{celestialBDS}

We would like to start by investigating the divergences of the celestial amplitude \eqref{MBDS}. To do so, let us expand the divergent part of the exponent in \eqref{BDS_TY} in powers of $\epsilon$
\begin{multline}
\label{qqq}
-\frac{1}{2}\sum_{l,i} a^l \left(\frac{\gamma^{(l)}}{4 (l \epsilon)^2} +  \frac{\cG^{(l)}}{2 l \epsilon} \right)\left(-\mu^2 s_{i,i+1}^{-1}\right)^{l \epsilon}
=-\frac{n\gamma_{2}}{8 \epsilon^2}  -\frac{n\cG_{1}}{4\epsilon} 
- \frac{ \gamma_{1}}{8\epsilon} \sum_i \log\left( -\mu^2 s_{i,i+1}^{-1}\right) 
\\ -\frac{1}{4}\cG \sum_i \log\left( -\mu^2 s_{i,i+1}^{-1}\right)-\frac{1}{16}\gamma \sum_i \log^2\left( -\mu^2 s_{i,i+1}^{-1}\right)+O(\epsilon)\,,
\end{multline}
where we have defined
\be
\label{eqgamma}
\gamma_m(a)=\sum^{\infty}_{l=1}  l^{-m} a^{l} \gamma^{(l)}\,, \quad \cG_m(a)=\sum^{\infty}_{l=1} l^{-m}a^{l}  \cG^{(l)}\,,
\ee
with the understanding that $\gamma_0(a)=\gamma(a)$ and $\cG_0(a)=\cG(a)$ as defined in \eqref{gaG}. To perform the Mellin integrals in \eqref{MBDS}, we first split the exponential in divergent and finite terms. We then promote the Mandelstam variables to operators acting on conformal primaries. As we will see below, the use of operators of this kind streamlines the expression for the hard celestial amplitude.  

Let $\hat P_i$ be an operator such that \cite{Arkani-Hamed:2020gyp}
\be
\label{opP}
\hat{P}_i \cO_{\omega_i}(p_i)  =\eta_i \omega_i  \cO_{\omega_{i}}(p_i) \quad \Longrightarrow \quad \hat{P}_i \cO_{\Delta_i} (z_i,\bar{z}_i) = \eta_i  \cO_{\Delta_i+1}  (z_i,\bar{z}_i)
\ee
where $\cO_{\omega_i}(p_i)$ is the $i$-th gluon state with momentum $p_i$, and $\cO_{\Delta_i}=\int^{\infty}_0 \frac{d\omega_i}{\omega_i} \omega_i^{\Delta_i} \cO_{\omega_i}$ is the corresponding primary field defined on the celestial sphere. Acting on a celestial amplitude, one can use $\hat{P}_i$ to define the Mandelstam operators
\be
\label{mandeop}
\hat{s}_{ij}= \eta_{i} \eta_{j} \hat{P}_i \hat{P}_j |z_{ij}|^2\,,
\ee
whose eigenvalues on plane waves are the Mandelstam variables. 

Let us now go back to the Mellin integral \eqref{MBDS}. From \eqref{qqq},  we see that the full celestial amplitude factorizes into a divergent part times a finite one,\footnote{We would like to stress that expression \eqref{celestial-factorization} coincides with previous findings for $n=4$ particles \cite{Gonzalez:2020tpi}. There, the Mellin integral was performed order-by-order in the loop expansion and, following the same logic as in \cite{Bern:2005iz} after resumming all contributions, it was shown to exponentiate yielding \eqref{celestial-factorization}. This has been worked out in appendix \ref{bdsa}.} \emph{i.e.},
\be
\label{celestial-factorization}
\tilde \cM_n = \tilde \cM_{\rm div}\, \tilde \cM_{\rm hard}\,,
\ee
where, using the cyclic property $\delta_{1,n+1} \equiv \delta_{1,1}=1$, we have organized the divergent part as
\be
\label{divergent-planar}
\tilde \cM_{\rm div} =  [\cJ(\epsilon,a)]^n  \exp \left[\frac{ \gamma_{1}}{8\epsilon}   \sum_{i<j} ( \delta_{i,j+1} +\delta_{i+1,j} ) \log \left( |z_{ij}|^{2} \right) \right]\,,
\ee
with the jet function $\cJ (\epsilon,a)= e^{i \pi \frac{\gamma_{1}}{8 \epsilon}} e^{-\frac{\gamma_{2}}{8 \epsilon^2}  -\frac{\cG_{1}}{4\epsilon} }$ being a non-dynamical factor depending on the coupling only. The finite contribution becomes an operator acting on the tree-level celestial amplitude $\tilde{A}_{\rm tree}$, namely
\begin{multline}
\label{finite-planar}
\tilde \cM_{\rm hard} =  (\mu^2)^{-n\frac{\gamma_{1}}{8 \epsilon}} \exp \left[-\frac{1}{4}\cG \sum_i \log\left( -\mu^2 \hat{s}_{i,i+1}^{-1}\right)\right. \\ \left. -\frac{1}{16}\gamma \sum_i \log^2\left( -\mu^2 \hat{s}_{i,i+1}^{-1}\right) + F_{n}(a,\hat{s}_{ij})+O(\epsilon)\right]  \tilde{A}_{\rm tree}\left(\left\{ \Delta'_{i} \right\}\right) \,.
\end{multline}
The hard amplitude in \eqref{finite-planar} has been defined in terms of  shifted conformal dimensions $\Delta_i'=\Delta_i + \frac{ \gamma_{1}}{4\epsilon}$. Note that even though the dimensionful factor $(\mu^2)^{-\frac{\gamma_{1}}{8 \epsilon}}$ blows up as $\epsilon\to 0^-$, the full object defined in \eqref{finite-planar} will be finite after using the explicit expression for $F_{n}(a,\hat{s}_{ij})$. In fact, the exponential operator appearing in \eqref{finite-planar} simply enters through the integrals, transforming the $\hat s_{i,i+1}$ operators into the Mandelstam variables $s_{i,i+1}$, thus allowing us to explicitly evaluate the full expression for the finite amplitude $\tilde \cM_{\rm hard}$. This will be the focus of subsection \ref{fin}. However, to properly define \eqref{finite-planar} as a celestial correlator in terms of dressed operators as in \eqref{dressed0}, we first need to examine in more detail the conformal structure of the divergent piece \eqref{divergent-planar}.

\subsection{Divergent part as a conformal correlator}
\label{divlargeN}
The first relevant property of \eqref{divergent-planar} is that it transforms as a $n$-point celestial correlator in the sense that it is covariant under SL(2,$\mathbb{C}$). From the transformation
\be
\label{sl2c}
z'_i = \frac{a z_i+b}{c z_i +d} \quad \Longrightarrow \quad z'_{ij}= \frac{z_{ij}}{(cz_i+d)(cz_j+d)} \,,
\ee
one obtains
\be
\label{covariance}
 \tilde{\cM}_{\rm div} \left(\left\{\frac{a z_i+b}{c z_i +d}\right\}\right)= \prod^{n}_{k=1}\left|c z_k +d\right|^{-\frac{\gamma_{1}}{2\epsilon}}\tilde{\cM}_{\rm div}\left(\{z_i\}\right)\,,
\ee
thus yielding that the divergent part of the celestial amplitude behaves as a conformal correlator composed of spin-zero primary fields $\cV_{\sigma}(z_{i},\bar{z}_i)$ with weights
\be
\label{qq2}
\sigma=-\frac{\gamma_1(a)}{8\epsilon}\,. 
\ee
Note that $\sigma$ is positive due to the positivity of $\gamma_{1}(a)$. This follows from the fact that the cusp anomalous dimension $\gamma$ is positive for arbitrary values of the Yang-Mills coupling $\alpha$, \cite{Basso:2009gh}\footnote{The positivity of the cusp anomalous dimension in QCD has been discussed in \cite{Belitsky:2003ys}}. Then, one first concludes that $\gamma_1(a)$ is a monotonically increasing function of the coupling, but since $\gamma_{1}(0)=0$, the announced positivity is ensured.

Another appealing feature is the possibility of finding an explicit representation for the operators $\cV_{\sigma_i}(z_{i},\bar{z}_i)$. For such purpose, we introduce scalar fields $\Phi^a(z,\bar{z})$ with an extra color index $a$ labelling its $m$ entries.  We define the two-point function of these scalars as
\be
\label{2pts}
\langle \Phi^a(z,\bar{z}) \Phi^b(w,\bar{w})\rangle=-\sigma \, \delta^{ab} \log |z-w|^2\,.
\ee
Let $\{\boe_{1},\cdots,\boe_{n}\}$ be a set of $n \leq m$ vectors in $\mathbb{R}^{m}$ satisfying the orthornomality condition 

$$\boe^a_i \boe^a_j \equiv \boe_i \cdot \boe_j= \delta_{i,j}\,.$$ 
We use this basis of vectors to represent the interactions between nearest neighbors appearing in the divergent part of the BDS amplitude. For each external particle located at the position $(z_k,\bar{z}_k)$, we define  vectors  $\bl_k^a=\boe^a_{k}-\boe^a_{k+1}$ with the identification $\boe_{n+1} \equiv \boe_1$. This is an over-complete set, as one can verify that the total sum vanishes, \emph{i. e.},
\be
\sum^n_{k=1} \bl^a_k =0\,.
\ee
This constraint suggests a conservation law. We will later identify the vectors ${\bl}^a_{k}$ with the generators ${\bf T}_i^a$ in the adjoint, representing the color charge of each of the external gluons. The expression above thus simply reflects color conservation. We will briefly expand on this in subsection \ref{LargeNQCD} where we will also identify the dimension $m$.

For the purpose of expressing the divergent amplitude as an explicit celestial correlator, it is useful to rewrite the sum in \eqref{divergent-planar} in terms of the inner product\footnote{This relation is reminiscent of the ${\rm SU}(m+1)$ Cartan matrix  $K_{ij}=2\delta_{i,j} -\delta_{i,j+1} -\delta_{i+1,j}$. 
In fact, one could consider that $m$  is the number of Cartan generators of ${\rm SU}(m+1)$, in which case $\Phi^a(z,\bar{z})$ can be associated with the free field representation of a two-dimensional CFT with an extended ${\rm SU}(m+1)$ internal symmetry \cite{Fateev:1987zh,Fateev:2007ab}. However, in the present case, the position $i=n+1$ is identified with $i=1$, thus we have  $\bl_1 \cdot \bl_{n}=-1$, while $K_{1n}=0$. Therefore,  the structure displayed by $\bl_i \cdot \bl_j$ is not exactly associated to a Cartan matrix.}
\be
\label{tt}
\bl_i \cdot \bl_j= \delta_{i,j}-\delta_{i,j+1} -\delta_{i+1,j} +\delta_{i+1,j+1}\,.
\ee
One then considers vertex operators defined by
\be
\label{cV}
\cV_{\bl_k}(z,\bar z)= \cJ(\epsilon,a) : \exp\left[i\, \bl_k \cdot \Phi (z,\bar{z}) \right]:\,.
\ee
Using the two-point function \eqref{2pts} and \eqref{tt}, one can show that the divergent part of the amplitude can be explicitly written as a correlator on the celestial sphere as
\be
 \tilde{\cM}_{\rm div}(\{z_i,\bar{z}_i\})= \langle \cV_{\bl_{1}}(z_1,\bar z_1) \cdots  \cV_{\bl_{n}}(z_n,\bar z_n)\rangle\,.
\ee
Furthermore, the importance of identifying the explicit representation \eqref{cV}, is that we are now able to construct dressed operators controlling the hard celestial amplitude following the discussion of section \eqref{infh-s}. In particular, due to the OPE
\be
\label{OPE1}
\cV_{\bl_i}(z,\bar{z}) \cV_{-\bl_i}(w,\bar{w}) \sim \frac{\cJ^2}{|z-w|^{4 \sigma}}+\cdots\,,
\ee
it is possible to invert relation \eqref{dressed0} to obtain the corresponding hard operator,
\be
\label{dressed}
\hat{\cO}_{\Delta_i,\ell_i}(w,\bar{w}) = \cJ^{-2}\lim_{z \to w} |z-w|^{4\sigma } \cV_{-\bl_i}(z,\bar{z})\, \cO_{\Delta_i,\ell_i}(w,\bar{w})\,,
\ee
where we can clearly see that the role of $\cV_{-\bl_i}$ is to shift the conformal dimension by $-\sigma_i$. Also notice that, except for the jet function, this expression coincides with previous results for celestial amplitudes in scalar QED \cite{Arkani-Hamed:2020gyp}. Thus, the hard BDS celestial amplitude is defined through
\be
\tilde{\cM}_{\rm hard} \equiv \langle \hat{\cO}_{\Delta_1,\ell_1}(z_1,\bar{z}_1)  \cdots \hat{\cO}_{\Delta_n,\ell_n}(z_n,\bar{z}_n) \rangle =\frac{\tilde{\cM}_{n}}{ \tilde{\cM}_{\rm div}}.
\ee
It is worth stressing here that, since $\cJ$ and the operators $\cV_{\bl_i}$ are regulator dependent, they are not measurable quantities. However, \emph{ratios} of divergent correlators, like the hard amplitude defined above, are finite and give rise to IR-safe observables. In the next subsection, we show that this expression precisely matches with the computation of the Mellin integral \eqref{finite-planar} for four and five gluons.

\subsection{Finite part: The UV and the positivity of the cusp anomalous dimension}
\label{fin}
We start by evaluating the finite part of celestial factorization in \eqref{finite-planar} for the case of four and five gluons. The fact that this is even possible is thanks to the main feature of the BDS formula, namely, that the infrared-finite part of the all-loop amplitude exponentiates as well. As a by-product, we will see how the positivity of the cusp anomalous dimension $\gamma(a)$ is directly responsible for the existence of the corresponding hard CFT$_2$ correlation function on the celestial sphere. 

\subsubsection{Four gluons}
Taking the Mellin transform on each external particle, the explicit form for the finite part of the four-gluon celestial amplitude is
\begin{multline}
\tilde \cM_{\rm hard} = \prod_{i=1}^4 \left(\int_0^\infty d\omega_i \, \omega_i^{\Delta_i'-1}\right) \, \exp \left[-\tfrac{1}{4}\cG \sum_i \log\left( -\mu^2 {s}_{i,i+1}^{-1}\right)-\tfrac{1}{16}\gamma \sum_i \log^2\left( -\mu^2 s_{i,i+1}^{-1}\right) \right]  \\
\mu^{-\frac{\gamma_1}{\epsilon}}\exp\left[  \tfrac{1}{8}\gamma(a) (\log^2(-r) + \tfrac{4}{3} \pi^2) +C(a)+O(\epsilon) \right] A_{\rm tree} \, , 
\end{multline}
Here, $r$ is the conformally invariant cross-ratio
\be
\label{r}
r=\frac{z_{12}z_{34}}{z_{23}z_{41}}\,,
\ee
and $C(a)$ is a function of the coupling only.\footnote{For the detailed expression, see for instance, \cite{Bern:2005iz}.} The integral above can be performed along the lines of \cite{Gonzalez:2020tpi} (see Section 2.2) where, for definiteness, we assume two ingoing ($\eta_1=\eta_2=-1$) and outgoing particles ($\eta_3=\eta_4=1$).  In terms of the shifted conformal weights 
\be
\label{qq1}
h_i'=h_i-\sigma\,, \quad \bar{h}_i'=\bar{h}_i-\sigma\,,
\ee 
we obtain a four-point CFT$_2$ correlation function with
\be
\label{correlator}
\tilde{\cM}_{\rm hard}(\{\Delta'_i,J_i,z_{ij}\})= f_{\Delta'_i}(r,\bar{r}) \prod_{i<j} z_{ij}^{\frac{h'}{3}-h'_i-h'_j}\bar{z}_{ij}^{\frac{\bar{h}'}{3}-\bar{h}'_i-\bar{h}'_j}\,,
\ee
where $f_{\Delta_i'}$ is a conformally invariant factor, having a distributional nature in the $r$ variable\footnote{The precise form of the prefactor $I_{\Delta_i'}(r,\bar r)$ is somewhat irrelevant for our purposes here, but it is given by $$I_{\Delta_i'}(r,\bar r)=2 g^2\delta(r-\bar r)  \Theta(r-1) r^{1+\frac{\Delta'}{6}+\frac{1}{2}\cG} (r-1)^{\frac{\Delta'}{6}}e^{\frac{\pi^2}{6} \gamma(a) +C(a)+O(\epsilon)}$$} , given by
\be
f_{\Delta_i'} =  I_{\Delta_i'}(r,\bar r) \mu^{-2\cG} e^{\frac{\gamma(a)}{8}\log^2(-r)}
\int^{\infty}_0 \frac{dw}{w} w^{\beta+\cG}
\exp \left[ -\frac{1}{8}\gamma \left( \log^2\left(-\frac{r w}{\mu^2} \right)+\log^2\left(\frac{w}{\mu^2} \right)\right) \right]\,,
\ee
with $\beta=\frac{1}{2}\sum_{i=1}^4(\Delta_i-1)+\frac{\gamma_1}{\epsilon}$. After the change $u=\log(w/\mu^2)$, one immediately sees that this is a Gaussian integral whose convergence in the UV depends crucially on the positivity of the cusp anomalous dimension $\gamma(a)$, a fact that has been shown to hold to all orders in perturbation theory  \cite{Belitsky:2003ys,Basso:2009gh}. Therefore, turning the argument around, we can venture to say that the very existence of the celestial amplitude, corresponding to the planar gluon amplitude in $\cN=4$ SYM, to all orders in the 't Hooft coupling, is directly related with the positivity of the cusp anomalous dimension. Evaluating the gaussian integral above yields
\be
\label{f-BDS}
f_{\Delta'_i}(r,\bar{r})=  I_{\Delta_i'}(r,\bar r) \mu^{2\beta} \sqrt{\frac{4\pi}{\gamma(a)}} \exp\left[ \frac{1}{\gamma(a)}\left(\beta+\cG(a) -\frac{\gamma(a)}{4} \log(-r)\right)^2\right]\,.
\ee
As a consistency check, with $\epsilon$ held fixed and taking the zero-coupling limit $a \to 0$, we should recover the tree level celestial amplitude. More precisely, we must obtain the same expression as in \eqref{correlator} but with bare weights $(h_i,\bar{h}_i)$ and with the conformally invariant function 
\be
\label{ftree}
 f_{\rm tree}(r,\bar{r}) =4\pi g^2  \delta(r-\bar{r}) \Theta(r-1)r^{\frac{5}{3}}(r-1)^{\frac{2}{3}} \delta(\beta)\,.
\ee
Indeed, note that \eqref{f-BDS} is a gaussian function since, in the $a\to 0$ limit with $\epsilon$ held fixed, $\beta$ becomes a pure imaginary number. From the fact that $\gamma(a)=\cO(a)$ and using the representation of the Dirac delta\footnote{In taking the limit, we have also used that $\cG(a)$ and $\gamma_1(a)$ are of $\cO(a)$.}
\be
\delta(x) \equiv \lim\limits_{a\to 0} \frac{1}{\sqrt{4\pi a}} e^{-\frac{x^2}{4 a}}\,,
\ee
one precisely recovers the tree level result \eqref{ftree}
\be
\lim_{a\to 0} f_{\Delta'_i}(r,\bar{r})= f_{\rm tree}(r,\bar{r})\,.
\ee
Since we now have an explicit expression for the full celestial correlation function corresponding to the all-loop four gluon (planar) amplitude in the bulk, we must make sure that, after computing its inverse Mellin transform, we arrive at the original BDS formula in momentum space. In particular, since this process involves integrating over $\beta$ from $-i\infty$ to $+i\infty$, the correlator better have a good behavior in the large $|\beta|$ region in order assure convergence. But since \eqref{f-BDS} is a gaussian function in $i\beta$, the inverse Mellin transform, indeed, converges rapidly.

\subsubsection{Five gluons}
In the four-particle case, because the scattering takes place on a plane, the four-dimensional Dirac delta enforcing total-momentum conservation yields an overall factor $\delta(r-\bar r)$ in front of the amplitude, \emph{i.e.}, the cross-ratio $r$ in \eqref{r} is forced to be real. The rest of the factors from the delta function localize three of the four energies $\omega_i$, yielding a single Mellin integration left.

For five-point celestial amplitudes, the 4D Dirac delta fully constrains four of the energies, which can be chosen to be $\omega_i^*$ for $i=1,\dots,4$. Then, one can write \cite{Fan:2020xjj},
\be
\delta^4\left(\sum_{i=1}^5 \eta_i \omega_i q_i\right) =  \frac{i}{4} \frac{(1-r_4)(1-\bar r_4)}{r_4-\bar r_4} \frac{1}{|z_{14}|^2 |z_{23}|^2} \prod_{i=1}^4 \delta(\omega_i-\omega_i^*)\,,
\ee
where $r_4$ is one of the two conformal invariant cross-ratios defined by\footnote{The relation between the cross-ratio $r$ of the four-gluon (previous subsection) and the $r_4$ ratio defined here, is $r_4=\frac{r}{r-1}$.}   
\be
r_i =\frac{z_{12}z_{3i}}{z_{13}z_{2i}}\,, \qquad i=4,5.
\ee
In the case of five gluons, the four energies $\omega_i$, $i=1, \dots, 4$, become localized at
\be
\label{omega*}
\omega_i^* = f_{i5} \omega_5 \,,
\ee
where the coefficients $f_{i5}$ are functions that depend only on $z_{ij}$ and the cross-ratios $r_i$.

The full expression for the 5-gluon celestial amplitude, for all values of the 't Hooft coupling, is
\begin{multline}
\label{cA5}
\tilde \cM^{\rm hard}_5 = \prod_{i=1}^5 \left(\frac{-\mu^2}{\eta_i \eta_{i+1} |z_{i,i+1}|^2}\right)^{-\frac{\cG}{4}} \frac{i}{4} \frac{(1-r_4)(1-\bar r_4)}{r_4-\bar r_4} \frac{1}{|z_{14}|^2 |z_{23}|^2} \\ 
\mu^{-\frac{5\gamma_1}{4\epsilon}} \prod_{i=1}^4 (\omega_i^*)^{\Delta_i + \frac{\gamma_1}{4\epsilon} +\frac{\cG}{2}-1} \int_0^\infty \, 
\frac{d\omega_5}{\omega_5} \omega_5^{\Delta_5+\frac{\gamma_1}{4\epsilon}+\frac{\cG}{2}} \exp\left\{-\frac{1}{16}\gamma(a) h_5 (\{\omega_i^*\}, \omega_5)\right\} A_5^{\rm tree}\,,
\end{multline}
where the function $h_5$, entering in the exponent above, is given by\footnote{Here we used the expressions in section 4.C of \cite{Bern:2005iz}.}
\begin{multline}
\label{h5}
h_5(\{\omega_i\})=\sum_{i=1}^5 \log^2\left(\frac{-\mu^2}{\eta_i \eta_{i+1} |z_{i,i+1}|^2 \omega_i \omega_{i+1}}\right)-15 \zeta_2\\
+\sum_{i=1}^5 \log\left(\frac{\eta_i \eta_{i+1} |z_{i,i+1}|^2}{\eta_{i+3}\eta_{i+4} |z_{i+3,i+4}|^2} \frac{\omega_i \omega_{i+1}}{\omega_{i+3}\omega_{i+4}}\right) \log\left(\frac{\eta_{i+1} |z_{i+1,i+2}|^2}{\eta_{i+3} |z_{i+3,i+4}|^2} \frac{\omega_{i+1}}{\omega_{i+3}}\right)\,.
\end{multline}
Before continuing, we would like to notice the following. All of the helicity structure of the original amplitude in momentum basis is contained in the tree-level factor $\cM_5^{\rm tree}$. This, together with the rest of the $|z_{i,j}|$ dependence in the prefactors above, nicely combine to produce the five-point CFT$_2$ correlation function, with the correct covariance under SL(2,$\mathbb{C}$) transformations. Therefore, since the energies $\omega_i^*$ \emph{do} transform under SL(2,$\mathbb{C}$), one may wonder whether the explicit dependence on the $\omega_i^*$ in the expression for $h_5$ in \eqref{h5} could spoil the aforementioned covariance. The answer is of course not, since from \eqref{omega*} and the explicit expressions for $f_{ij}$ (see Appendix A of \cite{Fan:2020xjj}) one notices that they transform in such a way that the final evaluated expression for $h_5 (\{\omega_i^*\}, \omega_5)$ ends up being conformal invariant.

After evaluating $h_5(\{\omega_i^*\}, \omega_5)$ one obtains
\eal{
h_5 = 20 \log^2(\tfrac{\omega_5}{\mu})+ b(r_i) \log(\tfrac{\omega_5}{\mu}) +c(r_i)\,,
}
where the coefficients $b(r_i)$ and $c(r_i)$ are functions of the conformally invariant cross-ratios only.
After making the change $u=\log(\omega_5/\mu)$, one sees that the integral in \eqref{cA5} is again a gaussian whose convergence is  assured by the positivity of $\gamma(a)$.

%%%%%%%%%%%%%%%%%%%%%%%%%%%%%%%%%%%%%%%%%%%%%%%%%%
\section{Infrared divergences in gluon amplitudes}
\label{Sec3}
%%%%%%%%%%%%%%%%%%%%%%%%%%%%%%%%%%%%%%%%%%%%%%%%%%

In order to extend the results of the previous section, we first describe the general structure of IR divergences of scattering amplitudes in SU$(N)$ gauge theory. Throughout this section we use the color-space formalism of Catani and Seymour \cite{Catani:1996jh,Catani:1998bh}. In this construction, contributions to amplitudes are vectors decomposed in the space of product of traces of $n$ generators $t^a$ in the fundamental representation of $\mathfrak{su}(N)$. 
Tree level $n$-gluon amplitudes are decomposed in terms of single traces. At loop level, apart from single traces, the decomposition also allows for the appearance of multi-trace contributions (for a review see e.~g.~\cite{Dixon:2011xs}).

It is convenient to introduce the color charge operator $\bT_i=\{T^a_i\}$, which is a vector associated with the $i$-th external gluon, with $N^2-1$ entries. It is defined in the adjoint representation, therefore, it acts as a commutator on a generator in the fundamental representation 
\be
\label{adaction}
\bT^a_{i} \, t^{b_i} \equiv -i f^{a b_{i} c} t^c = [t^{b_i},t^a]\,. 
\ee
Color charge is conserved in a scattering process. Therefore, when acting on amplitudes, the sum over all operators $\bT^a_i$ vanishes
\be
\label{color-conservation}
\sum^{n}_{i=1} \bT^a_i=0\,.
\ee
We shall need the inner product between color charges $\bT_{i}\cdot \bT_{j}=\bT^a_{i} \, \bT^a_{j}$. The Casimir operator $C_i=\bT^2_i$ is proportional to the identity and thus commutes with the color operators.

By using dimensional regularization in $D=4-2\epsilon$ dimensions, the all-loop $n$-gluon amplitude resummation exhibits the factorization property \cite{Becher:2009qa,Becher:2009cu,Gardi:2009qi,Gardi:2009zv},
\be
\label{AA}
\cM_n(\{p_i\})= {\bf Z}(\alpha(\mu^2),\left\{s_{ij}\right\})\, \cM_{\rm hard}(\{s_{ij}\})\,, \quad
{\bf Z} = \cP \exp\left\{-\frac{1}{2} \int^{\mu^2}_0 \frac{d \lambda^2}{\lambda^2} {\bf \Gamma}_n(\alpha(\lambda^2),\left\{s_{ij}\right\} )\right\}\,,
\ee
where the path-ordered exponential ${\bf Z}$ contains all the information about infrared divergences. The hard part of the amplitude, $\cM_{\rm hard}$ corresponds to a vector in color-space, and it is finite in the $\epsilon\to 0$ limit. The argument of the exponential in $\bZ$ is integrated up to an arbitrary energy scale $\mu$, and $\alpha(\lambda^2)$ is the coupling that is generated through the renormalization group equation
\be
\label{ren}
\frac{\d}{\d \log(\lambda^2)} \alpha(\lambda^2)= -\epsilon\,\alpha(\lambda^2)+\beta(\alpha)\,,
\ee
where $\beta(\alpha)$ is the beta function of the theory. We consider the soft anomalous dimension matrix ${\bf \Gamma}_n$ given by 
\be\label{anomalous} 
{\bf \Gamma}_n(\alpha,\left\{s_{ij}\right\} )= -\frac{1}{4} \hat{\gamma}(\alpha) \sum^{n}_{i=1} \sum^{n}_{j \neq i} {\bf T}_{i}\cdot {\bf T}_{j} \log\left( -\frac{s_{ij}}{\lambda^2} \right) -\sum^{n}_{i=1} \gamma_{J_i} (\alpha)+ {\bf \Delta}_n(\alpha, \left\{ \rho_{ijkl} \right\} )\,,
\ee
where the first term contains contributions from pairwise interactions. The proportionality constant $\hat{\gamma}(\alpha)=\gamma(\alpha)/C_A$ is the cusp anomalous dimension divided by the Casimir in the adjoint representation.\footnote{The scaling $\gamma= C \hat{\gamma}$, with $\hat{\gamma}$ independent of the Casimir, has been shown to hold up to three loops \cite{Moch:2004pa} and it is modified starting at four loops \cite{Henn:2019swt,Becher:2019avh}, where it has been computed including non-planar corrections for QCD and $\cN=4$ super Yang-Mills.} The second contribution in \eqref{anomalous} is a sum over the collinear anomalous dimensions associated with each gluon $\gamma_{J_i}$, and the last term ${\bf \Delta}_{n}$ is a non-vanishing contribution for $n\geq 4$ that starts at third order in the coupling $\alpha$ \cite{Almelid:2015jia},
\be
{\bf \Delta}_{n}= \sum^{\infty}_{l=3} \alpha^l {\bf \Delta}^{(l)}_n(\left\{ \rho_{ijkl} \right\} )\,.
\ee
The kinematic dependence of ${\bf \Delta}_n$ is only through the cross-ratios $\rho_{ijkl}$ 
\be
\rho_{ijkl}=\frac{s_{ij} s_{kl}}{s_{ik} s_{jl}}=\left|\frac{z_{ij} z_{kl}}{z_{ik} z_{jl}}\right|^2\,.
\ee
From the second relation above one immediately sees that the cross-ratios $\rho_{ijkl}$ are also invariant under SL$(2,\mathbb{C})$ transformations \cite{Almelid:2015jia,Almelid:2017qju}. Hence, it is a conformally invariant function on the celestial sphere. 

Notice that it was recently observed that \eqref{anomalous} receives extra contributions starting at four loops \cite{Becher:2019avh}. Unlike ${\bf \Delta}_n$, these extra terms are not constrained to depend on the cross-ratios $\rho_{ijkl}$ only, but rather depend explicitly on logarithms of the Mandelstam invariants $s_{ij}$. Fortunately, precisely thanks to this logarithmic behavior (and color conservation), we will see that the effect of including these four-loop corrections are under control in terms of the celestial correlation functions (See Section \ref{sec:remarks} for more details).

\subsection{Infrared divergences as a correlator of Wilson lines}
\label{Wilson}
The renormalization factor ${\bf Z}(\alpha, \{s_{ij}\})$ in \eqref{AA} can be expressed in terms of a correlator of the product of $n$ light-like Wilson lines of the form \cite{Gardi:2011yz,Gardi:2013ita}\footnote{In this expression we will actually need to introduce a damping factor in order to render the correlators finite. For a recent review on these developments see \cite{Milloy:2020hzi} and references therein.}
\be
\mathbfcal{W}_{p_i}(y,x)=\cP \exp\left( ig  \int^{x}_y d \tau \,  p_i \cdot {\bf A}(\tau p_i)  \right)\,,
\ee
where ${\bf A}_{\mu}=A^{a}_\mu \, \bT^a_i$ is the gluon field for each external particle with momentum $p_i$. In order to see this, we first note that the renormalization factor can be rewritten as
\be
\label{ZWilson}
{\bf Z}(\alpha(\mu^2),\left\{2\,p_{i}\cdot p_{j}\right\} )=\exp\left(-\frac{1}{4}\hat{K}(\alpha) \sum^n_{k=1} C_k \log\left(\tfrac{p_k^2}{\mu^2}\right) \right) {\bf Z}\left(\alpha, \left\{\mu^2 \gamma_{ij}\right\}\right)\,,
\ee
where 
\be
\label{gammaK}
\gamma_{ij}=\frac{2p_{i}\cdot p_{j}}{\sqrt{p_{i}^2} \sqrt{p_{j}^2}}\,, \quad \hat{K}(\alpha) = \frac{1}{2}\int^{\mu^2}_0 \frac{d \lambda^2}{\lambda^2} \hat{\gamma}(\alpha)\,.
\ee
Notice that both $\gamma_{ij}$ and the exponential prefactor in \eqref{ZWilson} are ill-defined expressions since $p_i^2=0$ for external gluons. Therefore, we shall think of them by first giving a small mass to each external gluon, and then taking the massless limit at the very end. When doing this, one obtains \cite{Becher:2009cu}
\be
\label{ZWilson2}
{\bf Z}(\alpha(\mu^2),\left\{s_{ij}\right\})=  \lim\limits_{p_k^2 \to 0} \exp\left(-\frac{1}{4}\hat{K}(\alpha) \sum^n_{k=1} C_k \log\left(\tfrac{p^2_k}{\mu^2}\right) \right)   \langle \mathbfcal{W}_{p_1}(0,\infty) \cdots  \mathbfcal{W}_{p_n}(0,\infty)  \rangle\,,
\ee
which is free of the aforementioned light-like divergences. However, notice that a different type of singularity will remain, namely, UV divergences in the Wilson lines arising from the presence of cusps. These infinities are in a one-to-one correspondence with the infrared structure of the full amplitude \eqref{AA}, and whose role on the celestial sphere will be the focus of the following subsection.

\subsection{Celestial gluon amplitudes}
\label{celestialQCD}
For the purpose of finding the celestial amplitude of \eqref{AA}, it is convenient to write the Mandelstam variables as in \eqref{mande} $s_{ij}=\eta_{i}\eta_j\omega_i\omega_j |z_{ij}|^2$. Using this and the color conservation condition \eqref{color-conservation}, one deduces that $\bGa_{n}$ decomposes as 
\be
\label{Gprop}
\bGa_{n}(\alpha,\left\{s_{ij}\right\})= \bGa_{n}\left(\alpha, \{\mu^2|z_{ij}|^2\}\right)+\frac{1}{2}\hat{\gamma}(\alpha) \sum^n_{k=1} C_k \log\left(\tfrac{\omega_k}{\mu}\right)\,.
\ee
The above separation of the angle-dependent part $z_{ij}$ and the energies of the external particles $\omega_k$ has an important consequence on celestial amplitudes. More precisely, it factorizes the renormalization factor as 
\be
\label{Zprop}
\bZ\left(\alpha, \{s_{ij}\}\right)= \bZ\left(\alpha, \{\mu^2|z_{ij}|^2\}\right)\prod^{n}_{k=1} \left(\tfrac{\omega_k}{\mu}\right) ^{-\frac{1}{2}\hat{K} C_{k}}\,, 
\ee
Then, taking the Mellin transform of the amplitude \eqref{AA}, we obtain 
\be
\label{MellinIR}
\tilde{\cM}_{n}(\left\{\Delta_{i},z_i,\bar{z}_i\right\})= \, \bZ\left(\alpha, \{\mu^2|z_{ij}|^2\}\right) \tilde{\cM}_{\rm hard}\left(\{\Delta_{i}-\tfrac{1}{2}C_i\hat{K} ,z_{i},\bar{z}_i\}\right)\,,
\ee
where $\tilde{\cM}_{\rm hard}$ is the Mellin transform of the hard amplitude $\cM_{\rm hard}$ with the $\mu^{\tfrac{1}{2}\hat{K} \sum_{k}C_k}$ prefactor from \eqref{Zprop} included in its definition. Notice that this is the same type of factorization obtained in the planar limit $\cN=4$ super-Yang-Mills in \eqref{celestial-factorization}.  However, this result is valid for any gauge theory (even without supersymmetry) because it follows from the general factorization displayed in \eqref{AA}.   

The distinctive property of the celestial correlator $\tilde \cM_n$, is that it affects the hard factor by a mere shift  in the conformal dimensions $\Delta_i$ of its primaries in the amount
\be
\label{shiftQCD}
\Delta_{i} \to \Delta_{i}-\frac{1}{2}C_i \hat{K} \,.
\ee
As the dimensional regulator $\epsilon$ is removed, $\hat{K}$ diverges, corresponding to an infinite shift, precisely matching with the one obtained for the case of planar $\cN=4$ SYM theory.

Recall that the factor $\bZ(\alpha,\mu^2|z_{ij}|^2)$ in \eqref{MellinIR} governs the infrared behavior of the full amplitude and it possesses the color structure of a SU$(N)$ tensor product of Wilson lines in the adjoint representation. Among its appealing features are that it is solely expressed in terms of the celestial data $z_{ij}$ and transforms covariantly under SL$(2,\mathbb{C})$. Furthermore, we can use the regulated version of \eqref{ZWilson2} with $p_{i}$ replaced by $\mu q_i$ to express this factor in terms of purely celestial variables as
\be
\label{corrZ}
\bZ\left(\alpha, \{\mu^2|z_{ij}|^2\}\right)=\lim\limits_{\delta \to 0} \exp\left(-\tfrac{1}{2}\hat{K}(\alpha) \log\left(\delta \right)  \sum^n_{k=1} C_k \right) \langle \mathbfcal{W}_{q_1}(0,\infty) \cdots \mathbfcal{W}_{q_n}(0,\infty)  \rangle\, , 
\ee
where $\mathbfcal{W}_{q_i}(0,\infty)$ is a semi-infinite Wilson line in the regulated direction $q_{i}\equiv q(z_i)$  given by
\be
q_{i}=\frac{1}{2}\eta_i \left(1+|z_i|^2+ \delta^2,z_i+\bar{z}_i, -i(z_i-\bar{z}_i), 1-|z_i|^2-\delta^2\right)\,,
\ee
satisfying $q_i^2=\delta^2$ with $\delta$ a dimensionless parameter. At this point, one may think that $\mathbfcal{W}_{q_i}$ can be regarded as a primary field on the celestial sphere. For this to be true, we would need to analyze the transformation properties of these Wilson line operators. At leading order in the regulator $\delta$, the action of SL$(2,\mathbb{C})$ on $q_i$ produces a simultaneous Lorentz transformation and a scaling, \emph{i.e.},
\be
q^{\mu}\left(\frac{a z_i + b}{c z_i +d }\right)= |c z_i + d|^{-2} \Lambda^{\mu}_{\,\,\,\nu} q^{\nu}(z_i)+O(\delta^2)\,.
\ee
On the other hand, since Wilson lines on the half real line do not transform under rescalings $q_i\to \lambda q_i$, for all $\lambda$, we conclude that these operators are SL$(2,\mathbb{C})$ invariant in the $\delta \to 0$ limit. Hence, they do not display the needed transformation laws accounting for the covariant properties of $\bZ(\alpha ,\mu^2|z_{ij}|^2)$ as a celestial correlator. However, due to the nature of the scattering process, we will see that $\mathbfcal{W}_{q_i}$ is a bare operator that suffers from UV divergences associated to the presence of cusp singularities. As such, its transformation properties after renormalization will become anomalous producing the desired covariance under SL$(2,\mathbb{C})$. 

Let us consider the simplest observable containing only two semi-infinite Wilson lines\footnote{In order to make the argument clearer, we have assumed that $\eta_{1}=-\eta_{2}=-1$.}
\be
{\bf W}(\alpha, \mu^2 \gamma_{12})=\langle \mathbfcal{W}_{q_1}(0,\infty) \mathbfcal{W}_{q_2}(\infty,0)\rangle\,.
\ee
Notice that the second operator in the equation above has the reversed order due to color charge conservation $\bT\equiv \bT_1=-\bT_2$. This fact allows us to interpret this correlator as a Wilson loop that closes smoothly at infinity, but has a cusp at the origin. This type of singularity produces the following renormalization group flow \cite{Polyakov:1980ca,Brandt:1981kf}
\be
\label{grW}
\frac{d \log({\bf W}_{\rm R})}{d \log(\lambda)}=-\Gamma_{\rm cusp}(\gamma_{12},\alpha(\lambda))\,,
\ee
where $\Gamma_{\rm cusp}$ is the anomalous dimension of the renormalized two-point function ${\bf W}_{\rm R}$ in the presence of a cusp singularity. To make contact with light-like Wilson lines, we study the properties of the  equation above in the $\delta \to 0$ limit. Remarkably, it has been found in \cite{Korchemsky:1987wg} that for all values of the YM coupling $\alpha$, the cusp anomalous dimension admits the following expansion 
\be
\label{Gll}
\Gamma_{\rm cusp}(\gamma,\alpha)= \frac{1}{2}C \hat{\gamma}(\alpha) \log\left(\tfrac{|z_{12}|^2}{\delta^2}\right)+O(1)\,,
\ee
with $C=\bT^2$ being the Casimir of either colored state, and where the $O(1)$ terms represent contributions that do not depend on the cusp angle, \emph{i.e.}, they are independent of the `distance' $|z_{12}|$ on the celestial sphere. This fact, together with the logarithmic behavior above, yields that under the global conformal transformations \eqref{sl2c}, the correlator ${\bf W}_{\rm R}$ \emph{does} transform covariantly. Therefore, we shall identify the renormalized version of $\mathbfcal{W}_{q_{i}}(0,\infty)$ as the primary operator of interest on a celestial CFT$_2$. As a consequence, its anomalous dimension becomes the conformal weight $\sigma_i$ of the spinless operator 
\be
\label{proposal}
\mathbfcal{V}_{\bT_{i}}(z,\bar{z}) \equiv \mathbfcal{W}^{\rm R}_{q_{i}}(0,\infty) \quad  {\rm with} \quad \sigma_i=\frac{1}{4}  C_i \hat{K} \,.
\ee
Summarizing, the correlator of two semi-infinite light-like Wilson lines can be obtained by integrating \eqref{grW} and using the limit \eqref{Gll}, 
\be
\langle \mathbfcal{V}_{\bT_{1}}(z_1,\bar{z}_1) \mathbfcal{V}_{\bT_2}(z_2,\bar{z}_2) \rangle =  \frac{\cJ^2}{|z_1-z_2|^{4 \sigma}} \delta_{\bT_1+\bT_2,0} \,,
\ee
where $\cJ=\cJ(\alpha,\epsilon)$ is, until now, an undetermined prefactor, while the Kronecker delta enforces color charge conservation. It is interesting to highlight the resemblance between the relation just found and the OPE involving operators $\cV_{\bl}$ in the case of planar SYM amplitudes \eqref{OPE1}. More precisely, an examination of the soft anomalous dimension \eqref{anomalous} in the case of two external gluons, \cite{Gardi:2009qi,Gardi:2009zv} reveals that the prefactor $\cJ$ is given by 
\be
\label{JetsQCD}
\log\left( \cJ(\alpha,\epsilon)\right)=\frac{1}{2}\int^{\mu^2}_0 \frac{d \lambda^2}{\lambda^2} \left[\gamma_{J} (\alpha) + \frac{1}{4} C \hat{\gamma}(\alpha) \log\left(-\frac{\lambda^2}{\mu^2}\right)\right]\,,
\ee
where each external gluon contributes with the same collinear dimension $\gamma_{J_i}=\gamma_{J}$. Interestingly enough, this quantity coincides with the jet factor defined for the first time in \eqref{divergent-planar} in the case of SYM planar amplitudes.

Having identified the Wilson line primary field as the responsible for the infrared divergences, one can extend the application of \eqref{dressed0} to the color space basis as
\be
\label{dressedQCD}
\mathbfcal{O}_{\Delta,\ell}(z,\bar z)=:\mathbfcal{V}_{\bT}(z,\bar{z}) \, \hat{\mathbfcal{O}}_{\Delta,\ell}(z,\bar z):\,,
\ee
that  enables us to properly define the colored celestial hard amplitude $\tilde{\cM}_{\rm hard}$ as a correlator of IR-safe operators $\hat{\mathbfcal{O}}_{\Delta,\ell}(z,\bar z)$.

In the next subsection we study the large $N$ expansion of \eqref{MellinIR}, arriving at some general results for gluon amplitudes in planar QCD. As a by-product, for the specific case of $\cN=4$ SYM, we will reproduce all of our findings already obtained in section 2.

\subsection{Large \texorpdfstring{$N$}{N} contributions}
\label{LargeNQCD}
We describe how the results obtained in section \ref{Sec2} can now be reproduced in the general framework of infrared divergences in non-abelian gauge theories. Let us first notice that, in the large $N$ limit, the color structures appearing in \eqref{anomalous} project the amplitude onto vectors in color space with single-traces only. In fact, the full amplitude in the planar limit can be reconstructed from $n$ (square roots of) Sudakov wedges between adjacent external particles \cite{Magnea:1990zb,Sterman:2002qn}. In particular, this argument implies that the higher loop order contributions to the soft anomalous dimension \eqref{anomalous}, must vanish, \emph{i.e.},
\be
{\bf \Delta}^{\rm planar}_{n}=0\,.
\ee
Thus, the only object that needs to be studied in more detail is the action of $\bT_i \cdot \bT_j $ on single traces. This has been analyzed in appendix B of \cite{Becher:2009qa}, and here we briefly review their argument. When acting on single traces of $n$ generators in the fundamental representation $t^{a_i}$, with fixed color structure, one has
\be
\label{TT}
\bT_i \cdot \bT_j \, {\rm tr}[t^{a_1}\cdots t^{a_n}]={\rm tr}[t^{a_1}\cdots [t^{a_i},t^a]\cdots[t^{a_j},t^a] \cdots t^{a_n}] \,.
\ee
Expressions such as the one above simplify greatly with the use of the identity\footnote{We use the normalization ${\rm tr}(t^a t^b)=\frac{1}{2}\delta^{ab}$.}
\be
\label{idtt}
t^{a} \, t^{b_1}\cdots t^{b_k} \,  t^{a}= \frac{1}{2}{\rm tr}[ t^{b_1}\cdots t^{b_k}] -\frac{1}{2N} t^{b_1}\cdots t^{b_k}   \,,
\ee
involving the product of $k$ generators. Then, one notices that the leading $N$ contributions in \eqref{TT} will only arise when the precise combination $t^a t^a$ is present or, in other words, when $i$ and $j$ are nearest neighbors. The operator $\bT_i \cdot \bT_j$ then effectively acts on single traces of $n$ generators, for $i \neq j$, as 
\be
\label{clave}
 \bT_{i}\cdot \bT_{j} \to -\frac{N}{2}(\delta_{j,i+1}+\delta_{i,j+1})  \,.
\ee
We can therefore conclude that, on the celestial sphere, infrared divergences in the large $N$ limit are fully encoded in the expression
\be
\label{qcd-largeN}
\bZ_{\rm planar}\left(\alpha, \{\mu^2|z_{ij}|^2\}\right)= \cJ^{n} \exp \left( 
-\frac{N\hat{K}(\mu^2)}{4} \sum_{i<j} (\delta_{j,i+1}+\delta_{i,j+1})\log(|z_{ij}|^2) \right)\,,
\ee
with $\hat{K}$ defined in \eqref{gammaK} and $\cJ$ given by relation \eqref{JetsQCD} where we have set $\gamma_{J_i}=\gamma_{J}$, which is true for external particles of the same kind, such a gluons. It is worth noticing that the expression for $\bZ_{\rm planar}$ applies to any non-abelian gauge theory in the large $N$ limit. Therefore, the discussion presented in section \ref{divlargeN} which allowed for the identification of the celestial operators $\cV_{\bl_i}$, also holds here provided one recognizes $\tfrac{1}{4}N\hat{K}$ as the conformal weight $\sigma$ of the primary field \eqref{cV}. Furthermore, the color operator $\bT_i$ is effectively replaced by the vector $\bl_i$ in $\mathbb{R}^{m}$. Since $\bT_i$ is in the adjoint representation, we notice the number of entries $m$ scales as $m \sim N^2$ in the large $N$ limit. 

Finally, the $\cN=4$ SYM case can be immediately recovered by recalling that beta function vanishes. Then, from \eqref{ren}, one has 
\be
\alpha(\lambda)= \alpha_0 \, \left(\frac{\lambda}{\mu}\right)^{-2\epsilon}\,,
\ee
with $\alpha_0$ is the coupling at the energy $\mu$. We then perform explicitly the integration over the energy scale $\lambda$ in \eqref{AA} for $\epsilon <0$. Using a series expansion in the coupling $\alpha$, the coefficients $\hat{K}$ and $\cJ$  are then given by
\be
N\hat{K}= -\frac{1}{2\epsilon} \gamma_{1}(a)\,, \quad \log(\cJ)= -\left(\frac{\gamma_{2}(a)}{8\epsilon^2}  + \frac{\cG_1(a)}{2\epsilon}  - i \pi \frac{\gamma_{1}(a)}{8\epsilon}   \right)\,,
\ee
with the identification $\cG^{(l)}= 2 \gamma^{(l)}_{J}$. We have also used that, in the planar limit, $N\hat{\gamma}_m(\alpha) \equiv \gamma_m(a)$ and $\cG_m(\alpha) \equiv \cG_m(a)$ with $\gamma_m$ and $\cG_m$ defined in \eqref{eqgamma}. Hence, the two quantities above show that infrared divergent contribution $\bZ_{\rm planar}$ exactly coincides with the factorization given in section \ref{celestialBDS}.

\section{Final remarks}\label{sec:remarks}

We have described the celestial representation of infrared corrected gluon amplitudes using the well-established results \cite{Bern:2005iz,Gardi:2009qi,Becher:2009cu,Becher:2019avh} in momentum basis. We have first analyzed the planar case and then we have discussed finite $N$ contributions, where the crucial object is the soft anomalous dimension matrix ${\bf \Gamma}_n$, \eqref{anomalous}. However, it has recently been found that this object gets corrected at fourth loop order \cite{Becher:2019avh}. Therefore, one may ponder on the precise effect this correction has on the celestial side. This has been emphasized in \cite{Magnea:2021tt} and here we would like to elaborate on it. 

The four-loop correction to ${\bf \Gamma}_n$ is \cite{Becher:2019avh}
\be
\Delta {\bf \Gamma}_n(\alpha(\lambda^2),\{s_{ij}\}) = - g(\alpha)\left[ \sum_{i\neq j} \left(\cD_{iijj} + 2 \cD_{iiij}\right) \log \left(-\frac{s_{ij}}{\lambda^2}\right)+  \sum_{i\neq j \neq k} \cD_{ijkk} \log\left(-\frac{s_{ij}}{\lambda^2}\right) \right]\,,
\ee
with 
\be
g(\alpha)= \cO(\alpha^4)\,, \quad \cD_{ijkl}=d_{abcd}\,{\rm tr}(\bT^a_i \bT^b_j \bT^c_k \bT^d_l)\,,
\ee
where $d_{abcd}$ is a symmetric invariant tensor given in terms of traces of symmetrized products of generators in the adjoint representation. That the kimematical dependence of this correction is logarithmic in the Mandelstam variables $s_{ij}$ is a very welcome feature, since it implies that its effect on the celestial CFT \eqref{MellinIR} consists in a correction to the conformal weight given by
\be
\sigma_{i}= \frac{1}{4}C_i \hat{K}^{\rm corrected}(\alpha)\,,
\ee
with (compare with \eqref{gammaK})
\be
\hat{K}^{\rm corrected}(\alpha)= \frac{1}{2} \int^{\mu^2}_0 \frac{d\lambda^2}{\lambda^2}\left(\hat{\gamma}(\alpha) + 4 g(\alpha)\frac{d^{abcd}d^{abcd}}{C_i N_i} \right)\,,
\ee
where $N_i$ is the dimension of the representation (for gluons, $N_i=N^2-1$). The second term inside the parentheses is precisely the four-loop correction to the cusp anomalous dimension found in \cite{Henn:2019swt,Becher:2019avh}. In this sense, our proposal for the celestial vertex operators \eqref{proposal} includes this correction after making the simple change 
$$\hat{K} \to \hat{K}^{\rm corrected}\,,$$ 
which shows the robustness of our results. Moreover, including these corrections, the analysis presented in this work can be straightforwardly extended to massless external fermions by simply using now the operator $\bT_i$ in the fundamental representation. 

\vspace{2mm}

There are other possible extensions of this work that would be interesting to pursue in more detail in the future. Here we summarize some of them.

One of the most interesting aspects of the BDS formula is the all-loop exponentiation of the finite part, a feature due to an extra symmetry enjoyed by the theory: dual conformal invariance. This symmetry is particularly manifest for the case of $n \geq 6$  particles, where the finite pieces kinematically depend, exclusively, on dual conformal cross-ratios. This is a consequence of the close connection between dual conformal symmetry and finite part of the BDS amplitude. It then seems worthwhile to investigate what dual conformal invariance can teach us about the deeper structures of the corresponding celestial CFT.  

\vspace{2mm}

Another interesting aspect has been emphasized in \cite{Himwich:2020rro} for graviton amplitudes. They have shown that there is a link between the resummation of IR divergences in graviton amplitudes and memory effects. Since these effects are a way to retrieve IR-safe information from the amplitude, it would be worth addressing this link in the present context under the light of the color memory \cite{Pate:2017vwa}.

\vspace{1cm}
\paragraph{Acknowledgements}
~\\[8pt]
We would like to thank Horatiu Nastase, Georgios Papathanasiou, Andrea Puhm, and Ryo Suzuki for discussions. We especially thank Lorenzo Magnea for useful comments on the first version of this article. The work of H.G. is funded by FONDECYT grants 11190427 and 1210635. H.G. would like to thank the support of Proyecto de cooperación internacional 2019/13231-7 FAPESP/ANID. The work of F.R. has been supported by FONDECYT grants 11171148 and 1211545.

\appendix
\section{Four-point celestial BDS formula}
\label{bdsa}
A representation of the BDS formula on the celestial sphere has been obtained in \cite{Gonzalez:2020tpi} for the scattering of four gluons. The expression corresponds to a differential operator acting on the tree level amplitude. It reads
\be
 \tilde{\cM}_4= \exp\left( \sum^{\infty}_{l=1} a^l \left(f^{(l)}(\epsilon) \cF_1(r,l\epsilon) + C^{(l)}+ \cE^{(l)}(\epsilon)\right) \hat{\cP}^{l\epsilon}\right) \tilde{A}_{\rm tree}
\ee
where
\begin{align}
 \cF_1(r,\epsilon)&=-\frac{1}{\epsilon^2}\left[1+ \left(-r\right)^{-\epsilon}\right] + \frac{1}{2} \log^2\left(-r\right) + \frac{2}{3} \pi^2 + O(\epsilon)\,,\\
 f^{(l)}(\epsilon)&=\frac{1}{4} \gamma^{(l)} +\frac{l}{2} \cG^{(l)} \epsilon + O(\epsilon^2)\,,
\end{align}
the operator $\cP$ is defined by
\be
\label{P}
\hat{\cP}= \mu^2 r^{\frac{1}{3}} (r-1)^{\frac{1}{3}} \prod_{i<j} |z_{ij}|^{-\frac{1}{3}} \prod^{4}_{k=1} \exp\left( \frac{i}{2} \frac{\d}{\d \lambda_k}\right)
\ee
 and $C^{(l)}$ are real numbers and $\cE^{(l)}(\epsilon)$ are non-iterating $O(\epsilon)$ functions of $r$. Using the definitions presented above, one obtains
 \be
  \label{g4}
 \tilde{\cM}_4= \exp\left(-\sum^{\infty}_{L=1} a^l  \left(\frac{\gamma^{(l)}}{4 l^2 \epsilon^2} + \frac{\cG^{(l)} }{2 l \epsilon} \right) \left[1+ \left(-r\right)^{-l\epsilon}\right] \hat{\cP}^{l\epsilon}\right) \tilde{\cH}_4\,,
 \ee
where 
\be
\label{eqSYMfinite}
\tilde{\cH}_4=\exp\left[  \tfrac{1}{8}\gamma(a) (\log^2(-r) + \tfrac{4}{3} \pi^2) +C(a)+O(\epsilon) \right] \tilde{A}_{\rm tree}\,,
\ee
is a finite remainder of the amplitude as $\epsilon \to 0$, while $\gamma(a)$ and $C(a)$ are expressed in terms of a series of positive powers of the 't Hooft coupling $a$.

The celestial amplitude \eqref{g4} can be rewritten in a similar fashion to the infrared divergent correlators analyzed in section \ref{celestialBDS}. In order to obtain such a representation, we use the identity
\be
\label{stu}
 (s t u)^2= \left(\prod_{i=1}^4 \omega_i \right)^3 \prod_{i<j} |z_{ij}|^2,
\ee
implying
\be
\label{top}
t=-r^{-\frac{1}{3}} (r-1)^{-\frac{1}{3}} \prod_{i<j} |z_{ij}|^{\frac{1}{3}} \left(\prod^{4}_{k=1} \omega_k\right)^{\frac{1}{2}}\,.
\ee
Then, we promote the Mandelstam variables to translation operators acting on primary fields. Comparing \eqref{P} with \eqref{top}, and using the action of the translation operator \eqref{opP}, one identifies that in the conformal basis
\be
\hat{P}=\exp\left(\frac{\d}{\d \Delta} \right),
\ee
and thus one can define the operator $\hat{t}= -\mu^2\,\hat{\cP}^{-1}$ and $\hat{s}= r \mu^2\,\hat{\cP}^{-1}$. In terms of these operators, the celestial four-point amplitude becomes
\begin{multline}
 \tilde{\cM}_4= \exp\left(-\sum^{\infty}_{l=1} a^l  \left(\frac{\gamma^{(l)}}{4 l^2 \epsilon^2} + \frac{\cG^{(l)} }{2 l \epsilon} \right) \left[\left(- \mu^2 \hat{t}^{-1}\right)^{l\epsilon}+ \left(-\mu^2 \hat{s}^{-1}\right)^{l\epsilon}\right] \right) \\\times \exp\left[  \frac{\gamma(a)}{8} (\log^2(-r) + \frac{4}{3} \pi^2) +C(a)+O(\epsilon) \right] \tilde{A}_{\rm tree}\,.
 \end{multline}
Recalling the operator $\hat s_{ij}= \eta_i \eta_j \hat P_i \hat P_j |z_{ij}|^2$ and using momentum conservation for two ingoing ($\eta_1=\eta_2=-1$) and two outgoing particles ($\eta_3=\eta_4=1$), we notice
 \be
 \hat{s}_{12} \tilde{\cM}= \hat{s}_{34} \tilde{\cM}\equiv \hat{s} \tilde{\cM}\,, \quad  \hat{s}_{14}  \tilde{\cM}= \hat{s}_{23} \tilde{\cM} \equiv -\hat{t} \tilde{\cM}\,,
 \ee
that permits us to find
\be
\label{BDShat}
\tilde{\cM}_4=\exp \left[ -\frac{1}{2}\sum^{\infty}_{l=1} a^l \left(\frac{\gamma^{(l)}}{4 (l \epsilon)^2} +  \frac{\cG^{(l)}}{2 l \epsilon} \right) \sum^{4}_{i=1} \left(-\mu^2 \hat{s}_{i,i+1}^{-1}\right)^{l \epsilon} \right] \times  \tilde{\cH}_{4}\,.
\ee
Therefore, upon acting with the Mellin transforms on each external state in \eqref{BDS_TY}, one arrives at the exact same expression \eqref{BDShat}.

\begin{footnotesize}
%\bibliography{N4}
\providecommand{\href}[2]{#2}\begingroup\raggedright\endgroup

\end{footnotesize}

\end{document}